\documentclass[fleqn,usenatbib]{mnras}

\usepackage{newtxtext,newtxmath}

\usepackage[T1]{fontenc}
\usepackage{ae,aecompl}

\usepackage{graphicx}	
\usepackage{amsmath}	
\usepackage{amssymb}	
\usepackage{xcolor}

\newcommand{\be}{\begin{equation}}
\newcommand{\ee}{\end{equation}}

\title[Spectra of secondary Cosmic Rays]{Effects of reacceleration and source grammage on secondary cosmic rays spectra}

\author[V. Bresci et al.]{
V. Bresci,$^{1}$\thanks{E-mail: bresci@arcetri.astro.it (VB)}
E. Amato,$^{1}$
P. Blasi,$^{2,3}$
G. Morlino$^{1}$
\\
$^{1}$INAF/Osservatorio Astrofisico di Arcetri, Largo E. Fermi 5 - 50125 Firenze, Italy\\
$^{2}$Gran Sasso Science Institute, Viale F. Crispi 7 - 67100 L' Aquila, Italy\\
$^{3}$INFN/Laboratori Nazionali del Gran Sasso, via G. Acitelli 22, Assergi (AQ), Italy
}

\date{Accepted XXX. Received YYY; in original form ZZZ}

\pubyear{2019}
\begin{document}
\label{firstpage}
\pagerange{\pageref{firstpage}--\pageref{lastpage}}
\maketitle
%
\begin{abstract}
The ratio between secondary and primary cosmic ray particles is the main source of information about cosmic ray propagation in the Galaxy. Primary cosmic rays are thought to be accelerated mainly in Supernova Remnant (SNR) shocks and then released in the interstellar medium (ISM). Here they produce secondary particles by occasional collisions with interstellar matter. As a result, the ratio between the fluxes of secondary and primary particles carries information about the amount of matter cosmic rays have encountered during their journey from their sources to Earth. Recent measurements by AMS-02 revealed an unexpected behaviour of two main secondary-to-primary ratios, the Boron-to-Carbon ratio and the anti-proton-to-proton ratio. In this work we discuss how such anomalies may reflect the action of two phenomena that are usually overlooked, namely the fact that some fraction of secondary particles can be produced within the acceleration region, and the non-negligible probability that secondary particles encounter an accelerator (and are reaccelerated) during propagation. Both effects must be taken into account in order to correctly extract information about CR transport from secondary-to-primary ratios.
\end{abstract}




\section{Introduction}
In cosmic ray (CR) physics, it is customary to classify particles based on whether they become energetic particles by direct acceleration of the interstellar plasma in CR sources, or they are born as energetic particles, in the interaction of cosmic rays with interstellar plasma: the former are termed primary particles, while the latter are termed secondary. The ratio of secondary to primary fluxes has then a crucial role in our understanding of the physical processes at work during the propagation of CRs from their sources to the Earth. Indeed it provides a direct estimate of the so-called grammage, namely the quantity of matter traversed by CRs in their journey through the Galaxy. Secondary to primary ratios at high enough energy are expected to be proportional to such grammage, which in turn is a decreasing function of energy per nucleon $E_k$, since it is proportional to $1/D(E_k) \sim E_k^{-\delta}$, where $D(E_k)$ is the CR diffusion coefficient in the Galaxy. 

Several spectral features that have emerged from recent observations carried out by PAMELA and AMS-02 appear to be in tension with this standard picture. The detection of breaks in the spectra of primary elements is best explained as a consequence of a change in the energy dependence of the diffusion coefficient \citep{Tomassetti,Blasi 2012,aloisio13, genolini17} rather than to proximity effects of some sources \citep{Thoudam2012} or subtle features of the acceleration process \citep{ptuskin}. This is also confirmed by the recent measurement of the $B/C$ ratio \citep{AMS02 secondaries} that shows a hardening at high energy. Even taking into account such slower dependence of $D(E_{k})$ on $E_{k}$ above $\sim 300$ GeV/n, some anomalous behaviours seem to persist: the $\bar{p}/p$ ratio is almost constant \citep{pbar} with energy, and the same is true for the $e^+/p$ ratio \citep{positrons}. The latter is clearly related to the well known issue of the rising positron fraction, which however is accommodated once the production of positrons in mature pulsars and bow shock nebulae is taken into account \citep{hooper,pulsars}. 

Nevertheless, the rather odd similarity between the positron, antiproton and proton spectra at the Earth has stimulated much discussion on radically new views of CR transport in the Galaxy \citep{Katz2010,Blum2013,lipari}. In these models, the basic issue under debate is whether the $B/C$ ratio can really be considered as a reliable indicator of the grammage traversed by CRs while propagating on Galactic scales. In principle some of the grammage could be accumulated in near-source regions and this would affect profoundly the general picture. For instance, \cite{cowsik} postulated the existence of dense cocoons around sources of CRs, where grammage is accumulated in an energy dependent manner, before escape into the Galaxy, where the grammage is assumed to be energy independent. \cite{cowsik} discussed how this would explain the flat $\bar p/p$ and $e^+/p$ ratios, while leaving the low energy $B/C$ ratio unchanged. At $E_{k}\gtrsim 100$ GeV/n a flattening of the $B/C$ should be observed. 

The issue of the physical interpretation of such cocoons remains open and is probably the most pressing problem encountered by this class of models (in addition to requiring steep injection spectra of protons, different source spectra for nuclei and leptons, and negligible energy losses for electrons and positrons \citep{lipari,Katz2010,Blum2013}). To our knowledge, the only possibility to realize the cocoons in nature is the onset of non-linear CR propagation effects. Prolonged confinement of CR particles in the source vicinity thanks to the self-generation of waves \citep{Malkov, dangelo1, Nava2016, Nava2019} can provide the necessary physical grounds, but whether the accumulated grammage is large enough to be important strongly depends on the ionization level of the local ISM surrounding the source. In any case, if a non-negligible fraction of the grammage is accumulated within small regions of space, these regions could make a prominent contribution to  the diffuse gamma-ray emission \citep{dangelo2}. 

We believe that these novel approaches to CR transport are very interesting and the role of non-linear effects near sources deserve a better understanding. On the other hand, it is also important to assess in a clear way whether the standard theory includes all the relevant contributions, so as to make sure that there is indeed a problem in describing the data. 
Here we concentrated on two such effects, that are typically ignored in the calculation of the secondary-to-primary ratios and that become important when looking for subtle effects: 1) shock reacceleration of secondary CRs at random encounters of a particle with a SN shock, and 2) spallation reactions that take place while primary particles are still within the acceleration region (source grammage). 

In what follows, we shall assume that the positron fraction is well described once the contribution of positrons from mature pulsars is included \citep{hooper,pulsars}. We recall that such models cannot account for antiprotons, hence the flat dependence of the $\bar{p}/p$ ratio requires an explanation. 


The importance of the first phenomenon depends on the probability that, during propagation, a CR encounters a particle accelerating shock: if this happens, the CR particle, being supra-thermal, will be automatically injected in the acceleration process. As discussed by \citet{Blasi 2017}, particle spectra that are harder than what the shock would provide, are left unaltered in shape by this process, and simply enhanced in terms of normalization. On the other hand, softer spectra than the shock would produce are flattened in slope, towards what the shock would provide. Secondary spectra, steeper at injection than those of primaries according to the standard paradigm, are clearly expected to be more affected by such a phenomenon, which can then partly account for the anomalies in the ratios.  

The second effect, namely the grammage accumulated by CR particles while still within their sources, was considered by \cite{aloisio13}, who showed that its inclusion in the computation of the propagated spectrum allows one to better reproduce the $B/C$ data both at high (TeV) and low energies \citep{VOYAGER}. In the present work we take into account this phenomenon through a more accurate and self-consistent treatment, showing that it is especially important for $\bar p$.

In order to describe the effects of these two phenomena we fix the diffusion coefficient in such a way that it provides a good description of the proton spectrum with injection of a power law in momentum. As discussed above, this can be achieved either by assuming a spatially dependent diffusion coefficient \citep{Tomassetti} or by accounting for the transition from self-generated waves to a Kolmogorov-like cascading \cite[]{Aloisio 2015}. Interestingly, the effect discussed there remains valid in a more complex scenario where the structure of the magnetic halo is generated self-consistently from the advection and cascade of magnetic turbulence generated in the disk \citep{Evoli-Halo2018}. We decided to adopt the diffusion coefficient provided by the calculations of \cite{Aloisio 2015}. This procedure is not completely self-consistent since in non-linear models the spectra of particles determine the diffusion coefficient and viceversa. However, since the main contributors to these non-linearities are protons and $He$ nuclei, in practice the decision of adopting the diffusion coefficient of \cite{Aloisio 2015} is justified based on the fact that the fit to the spectra of $H$ and $He$ observed by AMS-02 is common to that approach and the current one. 

We show that the inclusion of these two phenomena allows one to explain both the $B/C$ and $\bar p/p$ ratio satisfactorily within a {\it quasi}-standard description of CR acceleration and propagation, provided primary He nuclei are injected in the Galaxy with a slightly flatter spectrum with respect to both protons and heavier nuclei.

More specifically we assume that protons are injected with a (rigidity) spectrum $\propto R^{-4.2}$ and He nuclei $\propto R^{-4.12}$, while the injected spectrum of CNO nuclei is $\propto R^{-4.18}$, hence with a slope that is in between that of $p$ and $He$. Given these assumptions we show that the combination of reacceleration and grammage at the source allows to reasonably reproduce the spectra of $p$, $He$, $B$, $Li$, $C$, $N$, $O$ and $\bar p$.

The paper is organised as follows: in \S~\ref{sec:acceleration} we describe the reacceleration phenomenon at a shock front; in \S~\ref{sec:transport} we introduce reacceleration in the propagation of cosmic rays solving the transport equation first for primary and then for secondary particles, making clear for each of them the source term (\S~\ref{sec: src terms}); in \S~\ref{sec: results} we compare the results of our calculations of the fluxes of primary and secondary particles with available data; finally we draw our conclusions in \S~\ref{sec: conclusion}.

\section{Particle distribution function at the shock including re-acceleration}
\label{sec:acceleration}

The particle distribution function at a shock which is accelerating both freshly injected and pre-existing particles can be worked out by solving the advection-diffusion equation:
\begin{equation}
	\frac{\partial}{\partial z}\bigg[D(p)\frac{\partial f(z,p)}			{\partial z}\bigg]-u \frac{\partial f(z,p)}{\partial z}+\frac{1}	{3} \bigg(\frac{du}{dz}\bigg) \, p \, \frac{\partial f(z,p)}		{\partial p}+Q(z,p)=0 \, ,
	\label{eq:transp}
\end{equation}
where $f(z,p)$ is the distribution function of accelerated particles at position $z$ and momentum $p$, $D(p)$ is the diffusion coefficient (assumed to be independent of $z$ for simplicity) and $u$ is the velocity of the fluid upstream and downstream of the shock surface, assumed to be one-dimensional. The term $Q(z,p)$ represents the injection of fresh particles in the acceleration process and, setting the shock position as $z=0$, we can write it as:
\be
Q(z,p)=\eta \frac{n_1 u_1}{4\pi p_{\rm inj}^2} \, \delta(p-p_{\rm inj}) \, \delta(z)\ ,
\label{eq:qinj}
\ee
where $n_1$ and $u_1$ indicate the density and velocity of the fluid upstream of the shock, $\eta$ is the fraction of the incoming particle flux injected at the shock, and we assume that particles are all injected at the same momentum $p_{\rm inj}$. A non-linear theory of particle reacceleration was developed by \cite{blasi2004}.

If non-linear effects inducing the formation of a precursor upstream are neglected, the 1-D stationary equation (Eq.~\ref{eq:transp}) contains the main physical ingredients of the problem of acceleration and diffusion at the shock surface. The particle distribution function at the shock can be found by integrating Eq.~\ref{eq:transp} around the shock front and in the upstream, as illustrated {\it e.g.} by \citet{Blasi 2017} and references therein. One finds:
\begin{equation}
   f_0(p)=s \frac{\eta n_1}{4\pi p_{\rm inj}^3} \bigg( \frac{p}{p_{\rm inj}} \bigg)^{-s} + s \int_{p_0}^p \frac{dp'}{p'} \bigg(\frac{p'}		{p}\bigg)^s\, g(p') \,,
	\label{eq:f0}
\end{equation}
where $g(p)=f(-\infty,p)$ is the distribution function of pre-existing particles, called {\it seeds} hereafter, and $s=3r_{\rm gas}/(r_{\rm gas}-1)$ with $r_{\rm gas}=u_1/u_2$ the compression factor of the shock, which approaches 4 in the case of strong shocks. As long as the minimum momentum of seed particles, $p_0$, is low enough (below $\sim$ GeV) to make the integral term in Eq.~\ref{eq:f0} dominated by the upper limit, its choice has no practical implications and might differ from $p_{\rm inj}$. Whenever the spectra of seed primary or secondary nuclei are steeper than $p^{-s} (i.e. \sim p^{-4}$ for strong shocks) the reacceleration term returns a contribution that asymptotically approaches $\sim p^{-s}$. This is always the case in the Galaxy, at least at $E_k>10$ GeV/n, due to the diffusive propagation that steepens the injection spectra. As a result, in the case of primary nuclei, one expects that taking into account the effect of reacceleration will only change the normalization of the spectrum, while leaving the slope unaltered, and ultimately lead to a (slightly) revised estimate of the particle injection rate. On the other hand, in the case of secondary nuclei, the expectation is that reacceleration will lead to a flattening of the spectrum with respect to the case in which its effects are not included. 

On Galactic scales, the importance of this effect is modulated by the probability that a secondary product, such as a boron nucleus or an antiproton, encounters one or more shocks before escaping the Galaxy. We deal with this problem in \S \ref{sec:transport}.
%
%
%
\section{CR transport in the Galaxy}
\label{sec:transport}
We assume that interactions between Cosmic Rays and interstellar gas only occur within the galactic disk, of infinitesimal thickness $h_d$. Then, the transport equation for the distribution function $F_\alpha(z,p)$ of each species $\alpha$, in the direction perpendicular to the Galactic disk, $z$, can be written as:
\begin{eqnarray}
  -\frac{\partial}{\partial z} \left[ D_{\alpha}(p) \frac{\partial  F_{\alpha}}{\partial z}\right] 
  + v_A \frac{\partial F_{\alpha}}{\partial z} 
  + 2 h_d n_d v_\alpha \sigma_{\alpha}  F_{\alpha} \delta(z)  \hspace{1cm} \nonumber \\
  -\frac{2}{3}\, v_A \, p \, \frac{\partial F_{\alpha}}   {\partial p} \, \delta(z)
  + 2h_d \, \delta(z) \frac{1}{p^2}  \frac{\partial}{\partial p} 
  	\left[ p^2 \left( \frac{dp}{dt} \right)_{\rm Ion,\alpha} F_{\alpha} \right] = \nonumber \\
  = \left[Q_{\rm prim,\alpha}(p) + Q_{\rm reacc,\alpha}(p)+Q_{\rm sec,\alpha}(p')     \left(\frac{p'}{p} \right)^2 \, \frac{dp'}{dp} \right]                 \delta(z) \, . 
\label{eq:slab}
\end{eqnarray}
Let us describe the various terms. The first term in Eq.~\ref{eq:slab} represents particle diffusion due to magnetic perturbations. The second term is the advection due to Alfv\'en waves, whose speed depends on the strength of the magnetic field and on the gas density in the halo, i.e. $v_A=B_0/\sqrt{4 \pi m_p n_i}$. We introduced this term with a scenario in mind in which waves are generated by CRs themselves. In standard CR transport theory this term is often set to zero because it is assumed that the net wave velocity vanishes. This is the case if exactly the same number of waves are generated in all directions, which is not the case if the waves are self-excited. The remaining contributions on the LHS describe energy losses: fragmentation, adiabatic and ionization losses respectively. Spallation reactions, with cross section $\sigma_\alpha$, lead a nucleus to fragment in lighter elements. These are assumed to occur only in the disk since the density of the embedded gas ($n_d$) is there much larger than in the halo, where, however, CRs spend more time. This assumption is justified provided $n_{d}h/H\ll n_{H}$, where $H$ is the half-thickness of the halo and $n_{H}$ is the halo density. 

As far as adiabatic losses are concerned, the expression in Eq.~\ref{eq:slab} reflects our assumption that the advection velocity is constant in $z$ except for a sign inversion above and below the disk. On the RHS, we have introduced three different source terms: $Q_{\rm prim, \alpha}$, accounting for acceleration of primaries of species $\alpha$ from the thermal pool; $Q_{\rm reacc, \alpha}$ accounting for shock re-acceleration of both primaries and secondaries; $Q_{\rm sec, \alpha}$ accounting for the production of secondaries by the spallation reactions of primaries, occurring both within the sources and during propagation through the galactic disk. The expression adopted for the latter term clarifies that a nucleus $\alpha$ with momentum $p$ generally results from the spallation of a nucleus $\alpha'$ with momentum $p'$. The different contributions will be discussed in detail in the following sections.  

The quantity that CR experiments usually measure is the particle flux per unit kinetic energy per nucleon, $I_\alpha(z,E_k)$. Eq.~\ref{eq:slab} can be rewritten in terms of $I_\alpha(z,E_k)$, remembering that $p=A_\alpha \sqrt{E_k^2+2m_pc^2E_k}$, with $A_\alpha$ the mass number of nuclei of species $\alpha$, and $I_\alpha(z,E_k) dE_k= v_\alpha(p) F_\alpha(z,p) p^2 dp$. From the latter condition we derive:
\be
 I_\alpha(z,E_k)=A_\alpha p^2 F_\alpha(z,p)\ ,
 \label{eq:IF}
 \ee
 so that multiplying Eq.~\ref{eq:slab} by $A_\alpha p^2$, one obtains:
\begin{eqnarray}
    \label{eq:slabek}
  - \frac{\partial}{\partial z} \left[ D_{\alpha} \frac{\partial I_\alpha}{\partial z} \right] 
  + v_A \frac{\partial I_\alpha}{\partial z} 	
  + 2h_d n_d v_\alpha \sigma_{\alpha} I_{\alpha} \, \delta(z)- \hspace{1cm}  \\
  - \frac{2}{3} v_A \left[ \frac{\partial (p I_\alpha)}{\partial p} -3I_\alpha \right] \delta(z)
  + 2h_d \delta(z) \frac{\partial}{\partial p} \left[ \left(\frac{dp}{dt}\right)_{\rm ion,\alpha} I_{\alpha} \right] 
  = {\cal Q}_{\rm src} \delta(z)\ , \nonumber
\end{eqnarray}
where we have used the conservation of the particle Lorentz factor in spallation reactions, $p'=A_{\alpha'}\,p/A_\alpha$; we have rewritten the adiabatic losses through:
\begin{equation}
 \frac{2}{3} v_A A_\alpha p^3 \frac{\partial F_{\alpha}}{\partial p} 
 = \frac{2}{3} v_A \left[ \frac{\partial}{\partial p} (p I_\alpha)-3I_\alpha \right]  \,;
\end{equation}
we have used the relation $dp/dE_k=A_\alpha/v_\alpha$ to rewrite ionization losses; and we have defined:
\be
{\cal Q}_{\rm src}(p)= \left(Q_{\rm prim,\alpha}(p) + Q_{\rm reacc,\alpha}(p)\right)A_\alpha p^2+Q_{\rm sec,\alpha}(p')     A_\alpha p' \,^2  \, \frac{dp'}{dp}   \,.
\label{eq:calq}
\ee
Assuming that all particles escape from the Galaxy at a height $H$ above and below the Galactic disk (such that $I_\alpha(\pm H, p)=0$), the solution of Eq.~\ref{eq:slabek} in the region $z>0$ reads
\begin{equation}
  I_{\alpha}(z,E_k)=
  I_{\alpha,0}(E_k) \frac{1- \exp \left[-v_A(H-z)/D_\alpha \right]}{1-\exp (-v_AH/D_\alpha)} \, ,
\label{eq:sluz}
\end{equation}
being $I_{\alpha,0}(E_k) \equiv I_\alpha(z=0,E_k)$. Integrating Eq.~\ref{eq:slabek} between $z=0^-$ and $z=0^+$, with account of the fact that $(\partial I_\alpha/\partial z)_{0^-}=-(\partial I_\alpha/\partial z)_{0^+}$, one finds
\begin{eqnarray}
  - 2D_\alpha \bigg( \frac{\partial I_\alpha}{\partial z} \bigg)_{z=0}  
  + 2h_d n_d v_\alpha \sigma_{\alpha}I_{\alpha,0}
  + 2 v_A I_{\alpha,0}   - \frac{2}{3} v_A \frac{\partial (p I_{\alpha,0})}{\partial p} +  \\
  + 2h_d   \frac{\partial}{\partial p} \left[ \frac{A_\alpha}{v_\alpha} 
 	\left(\frac{dE_k}{dt}  \right)_{\rm Ion,\alpha} I_{\alpha,0} \right]
  = {\cal Q}_{\rm src}(p) \nonumber\,.
\label{eq:transphd}
\end{eqnarray} 
We now define the grammage for nuclei of type $\alpha$:
\begin{equation}
  X_\alpha=\frac{h_d n_d m v_\alpha}{v_A} \left(1-\exp\left[-v_AH/D_\alpha\right]\right) \, ,
  \label{eq:grammage}
\end{equation}
This represents the quantity of matter traversed by a CR. If the interstellar medium is made by $85\%$ of Hydrogen and $15\%$ of Helium, the average mass of gas particles is $m=1.4 m_p$. 

We then introduce the rate of adiabatic energy losses per unit grammage as \cite[see, e.g.][]{aloisio13}:
\begin{equation}
  \left( \frac{dE_k}{dX} \right)_{\rm Adv} = -\frac{2}{3} \frac{v_A}{2h_d n_d m } p
  = - \frac{2}{3} \frac{v_A}{2h_d n_d m c}  \sqrt{E_k^2+2m_pc^2E_k} \, ,
\end{equation}
and the energy loss rate of an ion of charge $Z_{\alpha}$ due to ionization of neutral atoms (only H and He are included) with ionization potential $\tilde{I}_s$:
\begin{eqnarray}
  \left( \frac{dE_k}{dX} \right)_{\rm Ion,\alpha} 
  = \frac{1}{ n_d m v_\alpha} \bigg( \frac{dE_k}{dt} \bigg)_{\rm Ion,\alpha}
  = \frac{1}{ n_d m v_\alpha\, A_\alpha} \left( \frac{dE}{dt} \right)_{\rm Ion,\alpha} \, = \nonumber \\
  = - \frac{2\pi r_e^2 c m_e c^2 Z_\alpha^2}{n_d m v_\alpha\, A_\alpha} \frac{1}{\beta} \sum_{n_{s=H,He}} 
  	n_s \left[ \ln{\frac{(2m_ec^2\beta^2 \gamma^2 )^2}{\tilde{I^2_s}(1+\frac{2 \gamma m_e}{m_p})}} -2\beta^2 \right] \,.
\end{eqnarray}
Following \cite{Strong 1998}, in the above expression we have neglected both the shell correction and the correction term for large $Z$ or small $\beta$.
Using these expressions, and recalling that $v_\alpha dp=A_\alpha dE_k$, we can rewrite Eq.~\ref{eq:transphd} as 
\be
  I_{\alpha,0}\left(\frac{1}{X_\alpha} + \frac{\sigma_\alpha}{m} \right)
  + \frac{d}{dE_k} \left\{  \left[ \left( \frac{dE_k}{dX} \right)_{\rm Adv} 
  	+ \left( \frac{dE_k}{dX} \right)_{\rm Ion,\alpha} \right] I_{\alpha,0} \right \} = {\cal Q'}_{\rm src} \nonumber  
\label{eq:transpfin)}
\ee
and finally recast it as 
\be
  \Lambda_{1,\alpha}(E_k)I_{\alpha,0}(E_k)
  + \Lambda_{2,\alpha}(E_k) \partial_{E_k} I_{\alpha,0}(E_k)={\cal Q'}_{\rm src}\ .
\label{eq:difftransp}
\ee
Imposing the boundary condition $I_{\alpha,0}(E_k \to +\infty )=0$, the solution of Eq.~\ref{eq:difftransp} reads:
\begin{eqnarray}
  I_{\alpha,0}(E_k) = \int_{E_k}^{\infty} dE_k'  
  			        \frac{{\cal Q'}_{\rm src}(E'_k)}{| \Lambda_{2,\alpha}(E'_k) |} \hspace{2cm}   \nonumber \\
  			        \times \exp \left[ - \int_{E_k}^{E'_k} dE_k'' 
			        \frac{\Lambda_{1,\alpha}(E''_k)}{| \Lambda_{2,\alpha}(E''_k) |} \right] \, .
\label{eq:sluI0ek}
\end{eqnarray}
The source term, ${\cal Q'}_{\rm src}$, will have a different form depending on whether one is dealing with primaries or secondaries, as discussed in the next section.

\section{Source terms for different species }
\label{sec: src terms}
\subsection{Re-acceleration term }
\label{sec: reacceleration term} 
Under the assumption that CRs are re-energized homogeneously in the Galactic disc, through crossing of SN shocks, the re-acceleration term (second term on the RHS of Eq.~\ref{eq:f0}) can be written as:

%
%
\begin{eqnarray}
A_\alpha p^2 Q_{\rm reacc,\alpha}(p)
= s \frac{V_{\rm SN} \mathcal R_{\rm SN}}{\pi R_d^2}  
 \int_{p_0}^p \left(\frac{p'}{p} \right)^{s-2} I_{\alpha,0}(E'_k) \frac{dp'}{p'}  \,, 
\label{eq:priminj}
\end{eqnarray}
where ${\cal R}_{\rm SN}$ is the rate of SN explosions in the Galaxy, $R_d$ is the radius of the galactic disk and $V_{\rm SN}$ is the volume of the remnant, as derived in detail in \S~\ref{sec:acceleration}. Finally, the distribution function of seed particles, $I_{\alpha,0}$, is the CR flux itself, evaluated at $z=0$. As a consequence, once the re-acceleration is included the solution has to be found by iteration, e.g. following the procedure described in \cite{Blasi 2017}. Since this process returns a contribution proportional to $p^{-s}$, exactly as direct acceleration, as far as primary spectra are concerned, its inclusion only impacts the estimate of the particle injection rate (contained in the normalization). On the other hand, it makes a fundamental difference for secondary nuclei, which, by definition, are not directly accelerated at shocks and hence lack a flat injection term.

The volume of the remnant is a key parameter of the problem. Indeed, the size of the remnant corresponds to a different age and thus to a different maximum energy achievable in the process of acceleration/reacceleration. During the Sedov-Taylor phase, the SNR radius evolves according to
\begin{equation}
 r_{\rm SN}(t)=r_{\rm ST} \left( \frac{t}{t_{\rm ST}} \right)^{2/5}\ ,
\end{equation}
where $r_{\rm ST}$ is the radius at the beginning of the adiabatic expansion, at age $t_{\rm ST}$ and we have assumed that expansion occurs in a constant density ISM. 
In order to estimate the effects of reacceleration, we must compute the average volume of SNRs in our Galaxy. We introduce the probability distribution $P\left(r_{\rm SN}\right)$, such that:
\begin{equation}
  P\left(r_{\rm SN}\right) dr_{\rm SN} = K_P \frac{dt(r_{ \rm SN})}{T_{\rm max}}\,,
\end{equation}
where $T_{\rm max}\approx 3 \times10^4\ {\rm yr}$ is the time for which a remnant is expected to behave as an active accelerator of high energy particles (before entering the radiative phase).

The proportionality constant $K_P$ is obtained from the condition that $\int_{r_{\rm ST}}^{r_{\rm Max}} P\left(r_{\rm SN}\right) dr_{\rm SN} =1$, which allows one to write the SNR distribution in radius as
\begin{equation}
  P\left(r_{\rm  SN}\right) = \frac{5}{2r_{\rm ST}} \left[\left( \frac{r_{\rm Max}}{r_{\rm ST}} \right)^{5/2} -1 \right]^{-1} 
  	\left(\frac{r_{ \rm SN}}{r_{\rm ST}} \right)^{3/2} \,.
\end{equation}
We can then compute the volume of the average SNR in the Galaxy as:
\begin{equation} \label{eq:VSNaverage}
  \bar{V}_{\rm SN} =\int_{r_{\rm ST}}^{r_{\rm Max}} P(r_{\rm SN})\frac{4}{3} \pi r^3_{\rm SN} dr_{\rm SN}
  	      = \frac{20}{33} \pi \frac{r^{11/2}_{\rm Max}-r^{11/2}_{\rm  ST}}{r^{5/2}_{\rm Max} - r^{5/2}_{\rm ST}} \,.
\end{equation}
Assuming $ 1\ M_\odot$ of ejecta, an explosion energy $E_{\rm SN}=10^{51}$ erg and an averag ISM density $n_i=1$ cm$^{-3}$, the typical radius of a SNR turns out to be: $\bar{r}_{\rm SN} = (3 \bar{V}_{\rm SN}/4\pi)^{1/3} \approx 12$ pc.

In addition to the size, the evolutionary stage of a SNR also impacts the maximum energy up to which it can accelerate particles, $E_{\rm max}$. We estimate $E_{\rm max}$ based on the assumption that it is determined by the growth of the non-resonant streaming instability \citep{Bell04, Schure and Bell 2014}. One finds \citep{Cardillo et al 15}:
\begin{equation}
  \ln\left(\frac{E_{\rm max}(t)}{m_pc^2}\right)E_{\rm max}(t) \approx \frac{\xi_{\rm CR} \sqrt{4\pi m_p n_i }\, e}{5  c} 
  	r_{\rm SN}(t) v_s^2(t)\,,
	\label{eq:emax}
\end{equation}
where $\xi_{\rm CR}\approx 10\%$ is the CR acceleration efficiency. Eq.~\ref{eq:emax} can be recast in a more general form (also suitable for particle species other than protons) by introducing the particle rigidity, $\rho_{\rm max}=p_{\rm max}/Z$, with $Z$ the particle atomic number. From Eq.~\ref{eq:emax} we find:
\begin{equation}
  \rho_{\rm max} \approx 100  Z \left (\frac{t}{t_{\rm ST}} \right)^{-4/5} \,{\rm TV}
  = 100  Z \left(\frac{r_{\rm SN}}  {r_{\rm ST}} \right)^{-2} \, {\rm TV} \,.
\label{eq:Rmax}
\end{equation}
The maximum energy would appear as a cut-off in the spectrum of re-accelerated particles. However, based on Eqs.~\ref{eq:VSNaverage} and \ref{eq:emax}, the average galactic remnant can provide $E_{\rm max}\approx 5$ TeV/n. Hence, below this energy, every SN encountered is efficient at re-energizing particles. In the energy range of our interest, namely $E<3-5$ TeV/n, we will then neglect the effects of the cut-off, having in mind that at higher energies it might contribute and modify the spectra.

\subsection{Source Grammage}
\label{sec: src grammage}

Interactions of accelerated particles inside the accelerator, before they escape to become CRs, lead to the generation of secondary products. The grammage accumulated by CRs inside the sources can be estimated to be at the level of $0.1-0.2$ g cm$^{-2}$ \citep{Aloisio 2015}, namely rather small compared to the grammage accumulated in the Galaxy at low energies. However, at high energies, where now we finally have measurements of the $B/C$ ratio, such correction may become appreciable. In fact, \cite{Aloisio 2015} estimated that a grammage of $0.17$ g cm$^{-2}$ allows us to get a better description of the B/C ratio at $E_{k}\gtrsim 100$ GeV/n.


In order to properly account for this contribution, that we refer to as source grammage, we introduce a source term for secondary nuclei which reads:
  
 \begin{eqnarray}
  \label{eq:secinj}
  A_{\alpha'} p'^2  Q_{\rm sec,\alpha} (p')= \hspace{4cm}  \\
  \sum_{\alpha' > \alpha} 2 \, h_d \, v_{\alpha'}(E_k)  \, \sigma_{\alpha' \alpha}  (E_k) \, n_{\rm src} \, N_{\rm src,\alpha'}(p')p'^2 A_{\alpha'}  \, ,\nonumber
  \end{eqnarray}
where $n_{\rm src} \approx 4n_d$ is the density of the gas located inside the source, $\sigma_{\alpha'\alpha}$ is the cross section for spallation of the bullet nucleus $\alpha'$ and $N_{\rm src,\alpha'}$ is the density of particles of species $\alpha'$ within the source. 
For the latter term one can write:
\begin{equation}
N_{\rm src,\alpha'}(p')= f_{0,\alpha}(p')\frac{\bar V_{\rm SN}}{2\pi R_d^2 h_d}T_{\rm SN} {\cal R}_{\rm SN}\, ,
\label{eq:nsrc}
\end{equation}  
with $f_{0,\alpha}(p)$ connected to the distribution function of accelerated primary CRs as given in Eq.~\ref{eq:f0}.
Restricting our calculations to TeV energies, particles are confined within the acceleration region for a time $T_{\rm SN}$ which is at least equal to the duration of the phase for which the remnant is an efficient accelerator. The latter stage is usually assumed to coincide with the Sedov-Taylor expansion phase, which ends at $T_{\rm max}$, when the remnant becomes radiative. Without considering non linear effects, that eventually extend the time for which CRs are trapped in the vicinity of their sources, we assumed $T_{\rm SN} = T_{\rm max}$. 

Following what just discussed, we include the source term resulting from production within sources also for antiprotons.
Antiprotons are secondary particles as well, but rather than from spallation reactions they result from inelastic interactions.
The main channel of $\bar p$ production is p-p scattering, but non-negligible contributions also come from p--He, He--p and He--He reactions \cite[see][for further details]{KDDM 2018}. Including all the mentioned interactions results in:

%

\begin{eqnarray} 
\label{eq:pbarinjsrc}
p' \, ^2 Q_{\bar p} =
\sum_{\alpha'=p,He}\sum_{j=p,He} 2 \, h_d \, v_{\bar{p}}  \, n_{{\rm src},j} \times \nonumber \\
\times
 \int_{E_{\rm th}}^{+ \infty} dE_{k,\alpha'}  \frac{d \sigma_{\alpha',j}}{dE_{k,\bar{p}}}A_{\alpha'} p'^2 N_{\rm src,\alpha'} (p') \, .
\end{eqnarray}

\subsection{Standard terms}
\subsubsection{Purely primary nuclei: \rm p, $^4$He, O}
Protons, Oxygen and $^4$He can be considered as purely primary nuclei in the sense that the contribution of the spallation of heavier elements to their flux is completely negligible. The source term appearing in Equation \ref{eq:calq} then reads:
\begin{eqnarray}
  A_\alpha p^2 \left[Q_{\rm prim,\alpha}(p)+ Q_{\rm reacc,\alpha}(p)\right]
  = A_\alpha p^2 f_0(p) \frac{V_{\rm SN} \mathcal R_{\rm SN}}{\pi R_d^2} \nonumber \\
 = s \frac{V_{\rm SN} \mathcal R_{\rm SN}}{\pi R_d^2}  
   \left[ K_\alpha A_\alpha \left(\frac{p}{p_{\rm inj}} \right)^{2-s} 
     		+ \int_{p_0}^p \left(\frac{p'}{p} \right)^{s-2} I_{\alpha,0}(E'_k) \frac{dp'}{p'} \right] \,,
\label{eq:priminj}
\end{eqnarray}
where $K_{\alpha}=\eta n_d/(4 \pi p_{inj})$ is a constant normalized to the observed flux.
Although the contribution to the spectral shape of primary nuclei as due to reacceleration is negligible, for completeness we retain the contribution of reacceleration also for primaries. 

\subsubsection{Mixed origin nuclei: \rm C, N}
Primary nuclei as Carbon and Nitrogen are directly injected in the ISM by SN explosions but a non-negligible fraction derives from heavier nuclei through spallation reactions. As secondary products, Nitrogen is predominantly created by fragmentation of Oxygen, while Carbon derives mainly from Oxygen and Nitrogen itself. The source terms related to spallation reactions occurring during the propagation in the Galaxy are written as:
 \begin{align}
&A_{N} p'^2  Q_{\rm sec, N} (p')  \frac{dp'}{dp} \nonumber= A_{N} p'^2  Q_{\rm sec, N} (p')  \frac{A_{\alpha'}}{A_N} = \\ 
&=p'^2 2h_d n_d  v_{\alpha'} \sigma_{\rm \alpha' N}  F_{0, \alpha'} (p) A_{\alpha'}=
2h_d n_d  v_{\rm O} \sigma_{\rm ON}  (E_k) I_{0,\rm O}  (E_k)
\end{align}
\begin{align}
&A_{C} p'^2  Q_{\rm sec, C} (p') \frac{dp'}{dp}= \sum_{\alpha'= \rm O, N}A_{C} p'^2  Q_{\rm sec, C} (p')  \frac{A_{\alpha'}}{A_C}= \nonumber \\
&=2h_d n_d \left[ v_{\rm O} \sigma_{\rm OC}  (E_k) I_{0,\rm O}(E_k)+ v_{\rm N} \sigma_{\rm NC}  (E_k) I_{0,\rm N}(E_k) \right] .
\end{align}
For these nuclei, the RHS of Eq.~\ref{eq:slabek} is overall the sum of the injection and reacceleration term discussed in the previous section and the contributions above. 
\subsubsection{Purely secondary nuclei: $\rm \bar{p}$, \rm $^3$He, B, Li}
\label{subsec: pure sec}

Secondary CRs such as $\rm \bar{p}$, $^3$He, B and Li only originate from interactions involving primaries and may be reaccelerated through occasional crossings of SN shocks.  As a consequence, the source term ${\cal Q}_{\rm src}$ for these species contains the reacceleration term and the grammage accumulated by primaries inside the accelerator, as derived in the dedicated sections \ref{sec: reacceleration term} and \ref{sec: src grammage} respectively, plus the standard production from interactions within the Galaxy which, for Boron, Helium and antiprotons, reads as: 
 \begin{equation}
\label{eq:secBinj}
A_{B} p'^2  Q_{\rm sec, B} (p')  \frac{dp'}{dp}=2h_d n_d  \sum_{\alpha'= \rm C, N, O} v_{\alpha'} \sigma_{\rm \alpha' B}  (E_k) I_{0,\alpha'} (E_k) \, ,
\end{equation}
\begin{equation}
A_{\rm ^3He} p'^2  Q_{\rm sec, ^3He} (p')  \frac{dp'}{dp}= 
2h_d n_d  v_{\rm He} \sigma_{\rm He \, ^3He}  (E_k) I_{0,\rm He} (E_k)\, ,
\end{equation}
\begin{equation} 
\label{eq:pbarinj}
p' \, ^2 Q_{\bar p} =
\sum_{\alpha'=p,He}\sum_{j=p,He} 
2 h_d   v_{\bar{p}} n_{d,j}  \int_{E_{\rm th}}^{+ \infty} dE'_{k,\alpha'} I_{\alpha',0} 
	\frac{d \sigma_{\alpha',j}}{dE_{k,\bar{p}}} \,.
\end{equation}
Notice that the corresponding term for Li is identical to Eq.~\ref{eq:secBinj} but requires an additional contribution resulting from the spallation of B, namely $v_{\rm B}\sigma_{\rm BLi}  (E_k) I_{0,\rm B}(E_k)$.
B and Li nuclei are mainly produced by primary C, N, O, but a small contribution also comes from spallation of Ne, Mg and Si which can be evaluated as an additional flux of $10 \%$ for B and $17\%$ for Li \cite[see][Fig. 7]{DRAGON18}. Moreover, a relevant contribution results from intermediate short lived nuclei and can account for up to a few tens percent to the B source term \cite[see][for further details]{DRAGON18}. 
Concerning the total Helium flux and antiprotons, the contribution from $^3$He is added to the primary $^4$He since AMS-02 cannot distinguish among the two, and we have considered for $\bar{p}$--production solely the channels which involve $p$, $^4$He, $^3$He, as before.
%
\section{Results}
\label{sec: results}

In this section we illustrate the main results of our calculations. CR fluxes are evaluated using Equation (\ref{eq:sluI0ek}) with specific injection and source terms depending on the species, as explained in \S~\ref{sec: src terms}. Concerning the total spallation cross section we used the following empirical formula, valid for Lithium and heavier elements:  
\begin{equation} \label{eq: spallation cross sections}
\begin{split}
  \sigma_{\alpha} =& 45 \, A_\alpha^{0.7} \left[1+0.016 \, \sin\left(5.3-2.63 \ln{A_\alpha}\right) \right] \\
    &\times \left[1-0.62 \,  e^{-E_k/200} \, \sin{(10.9E_k^{-0.28})} \right] \, \text{mb},
\end{split}
\end{equation}
where the kinetic energy per nucleon, $E_k$, is expressed in MeV. An overall multiplicative factor of $0.8$ must be used for Helium nuclei \citep{Letaw spallation cross sections}. The cross sections for secondary production are taken from \cite{DRAGON18,Evoli}, 
where a detailed description, including decay chains, has been developed by fitting empirical or semi-empirical existing functional forms to large samples of measurements from $100$ MeV to $100$ GeV.
For B production, the contribution from intermediate short lived nuclei is estimated to be around $\sim 35 \%$ \cite[see again Fig. 7][]{DRAGON18}. As a consequence, the latter cannot be neglected and it is directly accounted for in the cross sections for B-production from C,N,O nuclei.

The cross sections for antiproton production are instead taken from the tables in the supplemental material of \citet{KDDM 2018}. In particular we considered the following interactions: p--p, p--$^4$He, $^4$He--p, $^4$He--$^4$He, $^3$He--p  and  $^3$He--$^4$He, where the first element refers to the CR nuclei and the second to the ISM component. It is worth stressing that those cross sections have a dependence on energy stronger than the spallation cross sections in Eq. \ref{eq: spallation cross sections}, which are instead roughly constant above $\sim 1$ GeV/n.

\begin{figure}
	\includegraphics[width=0.5\textwidth]{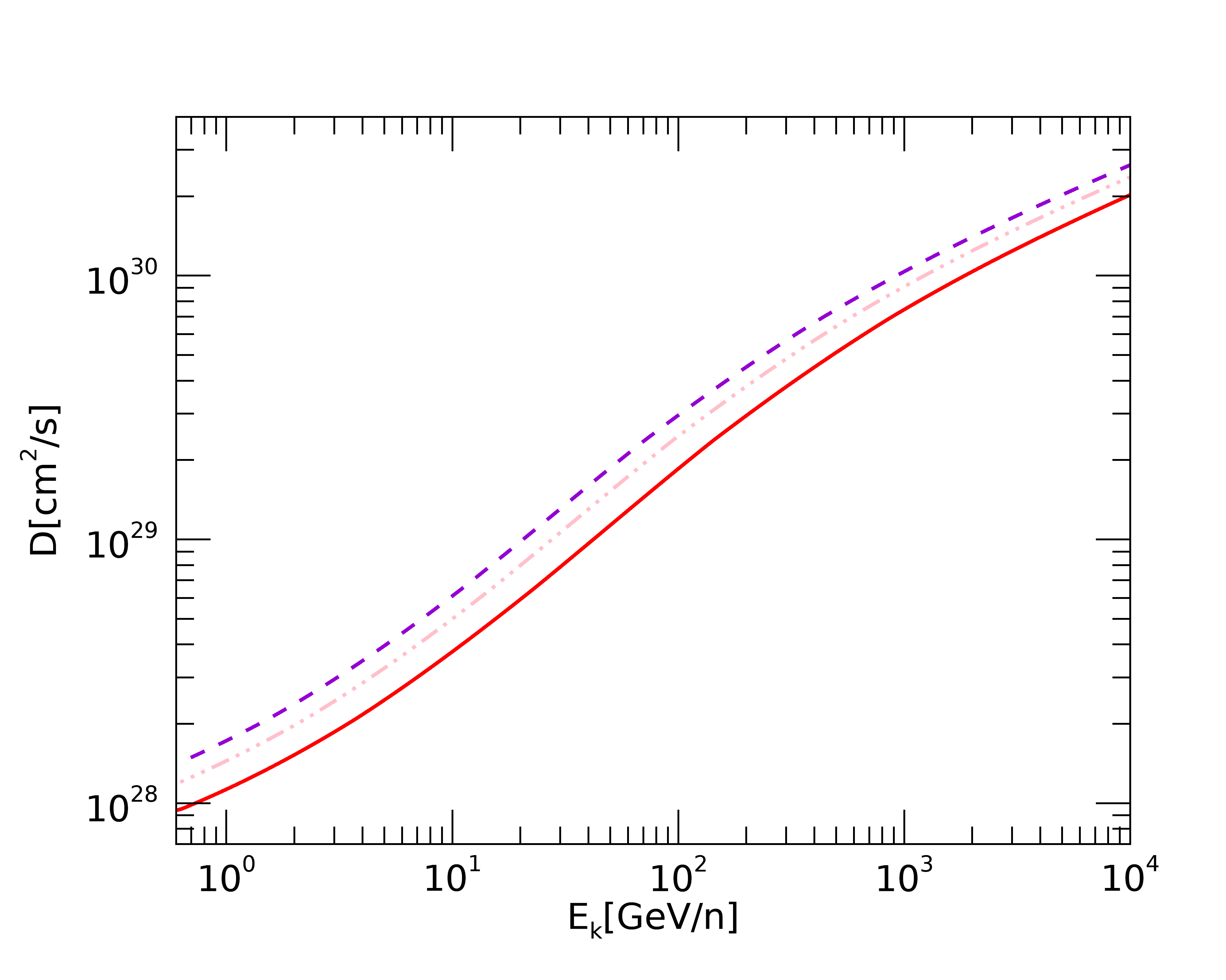}	
	\caption{Diffusion coefficient derived by calculations of \citet{Aloisio 2015} plotted against kinetic energy per nucleon. The dark violet dashed line, the pink dot-dashed one and the red continue curve represent the diffusion coefficient of nuclei with $A/Z \sim 2$, $\rm ^3He$ and protons respectively.} 
\label{Fig: Diffusion coefficient}
\end{figure}

As discussed above, in order to illustrate more effectively the role of reacceleration and source grammage, we decided to fix the diffusion coefficient to the one derived through the non-linear calculations of  \citet{Aloisio 2015}, where two contributions were taken into account: {\it i)} the Kolmogorov cascade of the large scale turbulence injected by (presumably) the same SNRs and {\it ii)} the self-generated turbulence produced by the streaming of CRs. While the former is relevant to scatter the highest energy CRs, the latter is more important at lower energies (see Figure~\ref{Fig: Diffusion coefficient}). A self consistent calculation shows that the transition between these two regimes occurs at rigidities around a few hundred GV: for $\sim 10\lesssim R\lesssim 300$ GV the slope of the diffusion coefficient is $\sim 0.6-0.7$, while at higher energies it asymptotically tends to the Kolmogorov-like spectrum ($\sim 1/3$). As a consequence, all CR spectra will naturally show a smooth break at a rigidity $\sim 300$ GV, as observed. Below $\sim 10$ GV, advection becomes important, as pointed out by \citet{Aloisio 2015}.

Since AMS-02 cannot discriminate among isotopes of a given element, our calculations are carried out for mean atomic mass number $A_B=10.8$ for B and $A_N=14.5$ for N. Figures from 2 to 8 show our main results. We carry out our calculations with and without the effect of reacceleration (Models 2 and 1 respectively). For each of the two cases we also show the results with or without accounting for the source grammage (labelled with subscripts {\it b} and {\it a} respectively). Table \ref{tab:mod} summarises all the models considered. The effect of solar modulation is accounted for by using the force-field approximation \citep{Gleeson solar mod}, with a potential $\phi_{\rm Sol} = 400$ MV. With the exception of $\bar{p}$, the unmodulated spectrum is also shown for the Model 2b (labelled as 2{\it bs}) and compared with Voyager data at low energies.

\begin{table}
\label{tab:mod}
\caption{Summary of all models considered.}
\begin{center}
\begin{tabular}{l|c|c|c}
\hline
                              \textbf{Model}           & \textbf{Reaccel.} & \textbf{Source Gram.}  & \textbf{Solar Mod.} \\
\hline
\textbf{1a:} 		& No                                            & No                                          & Yes                                  \\ 
\textbf{1b:} 		& No                                            & Yes                                         & Yes                                 \\ 
\textbf{2a:}              & Yes                                           & No                                          & Yes                                   \\ 
\textbf{2b:}              & Yes                                           & Yes                                         & Yes                                  \\
\textbf{2bs:}            & Yes                                            & Yes                                         & No                                 \\ 
\hline
\end{tabular}
\end{center}
\end{table}

\begin{figure}
	\includegraphics[width=0.52\textwidth]{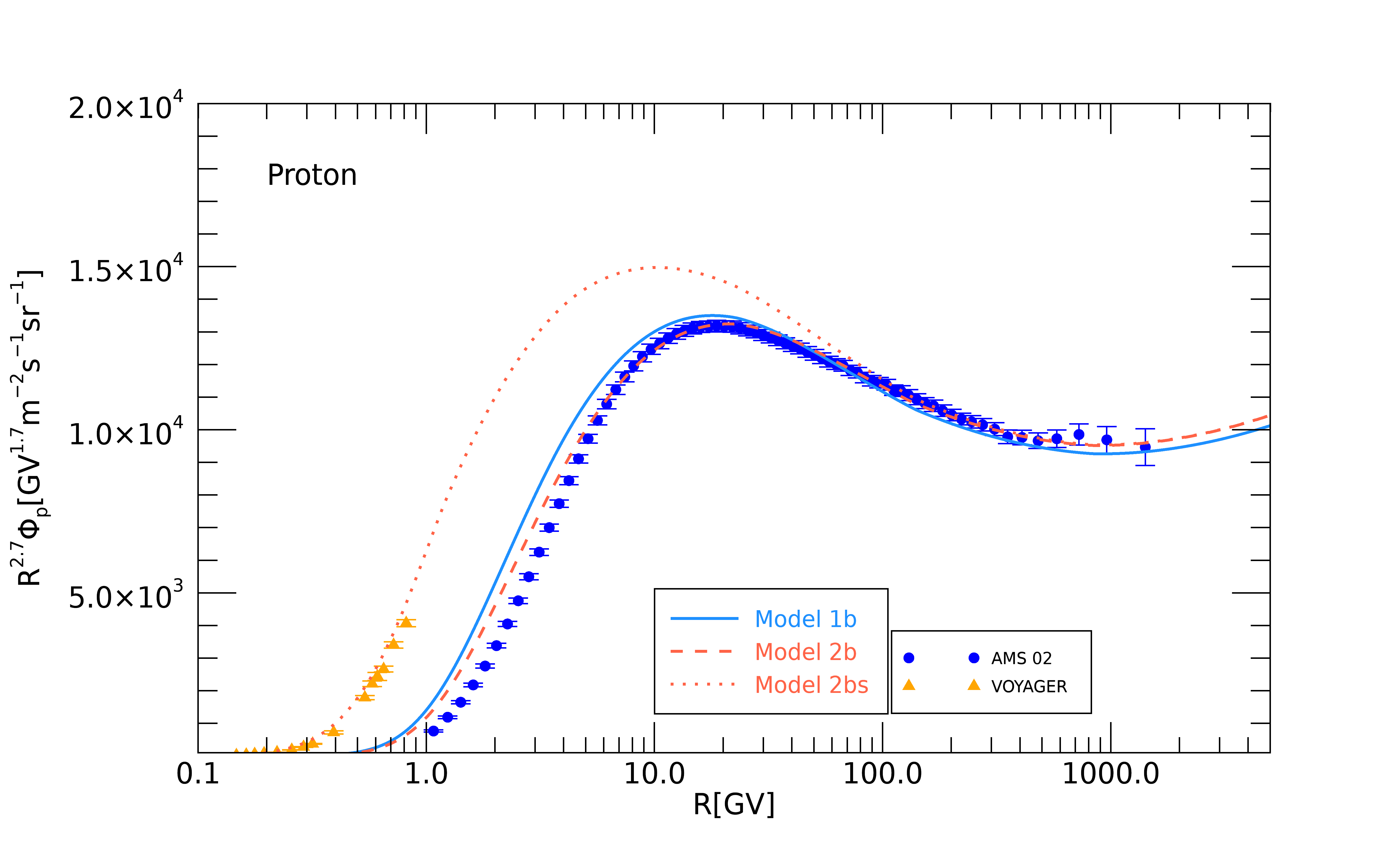}
	\caption{Spectrum of Protons: the blue points are the AMS-02 data \citep{AMS02 Protons}, and the orange triangles are from Voyager \citep{VOYAGER}. The curves illustrate our results without (Model 1b) and with inclusion of reacceleration (Model 2b). In all plots dot-dashed lines represent the latter model without account of solar modulation (Model 2bs), for direct comparison with Voyager data. See also Table~\ref{tab:mod}.}
	\label{Fig 1: Proton}
\end{figure}

In Figures \ref{Fig 1: Proton}, \ref{Fig: Carbon}, \ref{Fig: Oxygen} and \ref{Fig: Nitrogen} we show the spectrum of protons, Carbon, Oxygen and Nitrogen nuclei, respectively. The best agreement with the data is obtained assuming an injection slope $s=4.2$ for protons, $s=4.12$ for Helium and $s=4.18$ for CNO. It is worth stressing that while the difference in the slope between proton and heavier nuclei could, in principle, be due to uncertainties in the cross sections, the difference between proton and Helium, being larger, is difficult to explain as a consequence of the cross section uncertainties alone.

\begin{figure}
	\includegraphics[width=0.5\textwidth]{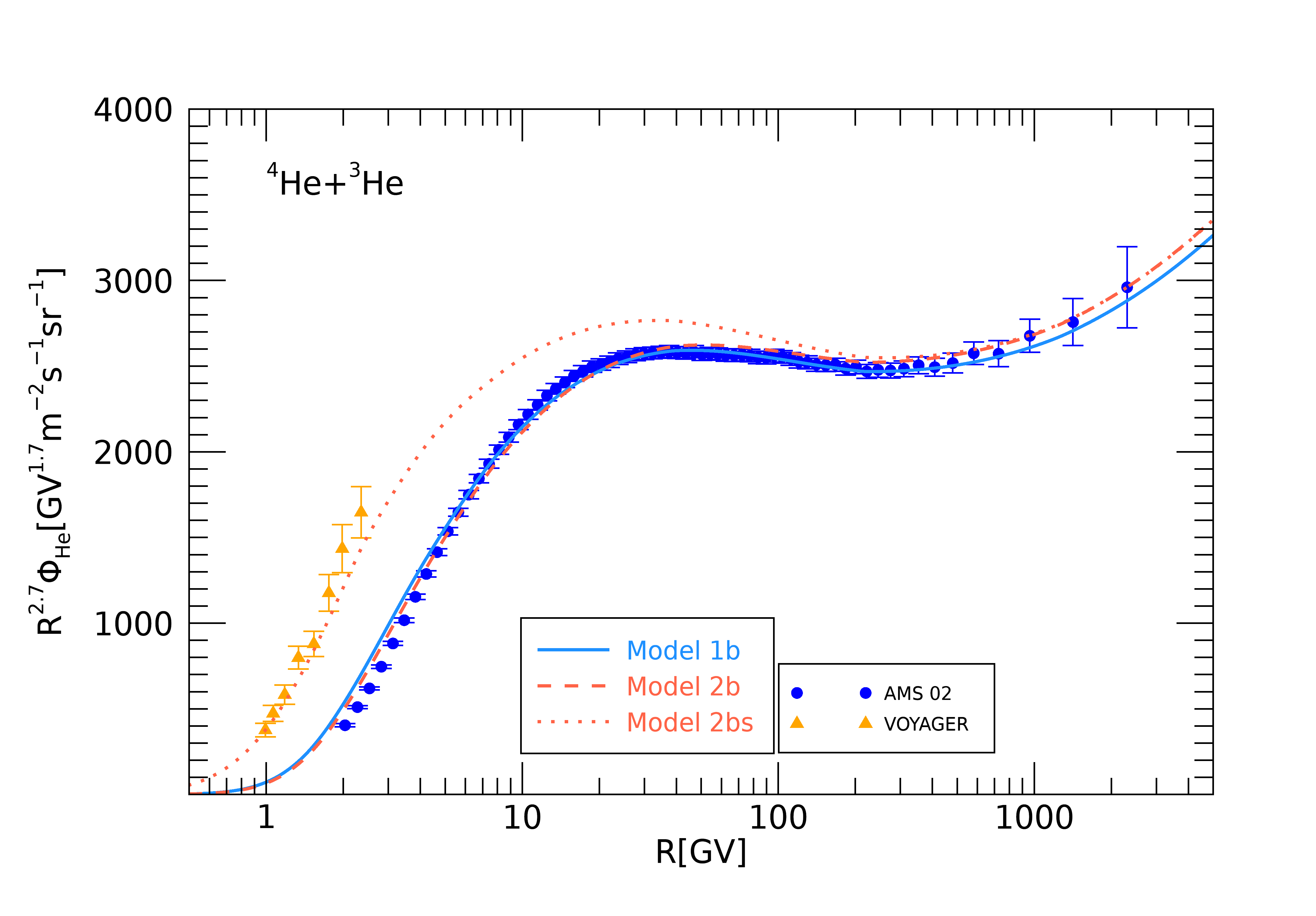}	
	\caption{Spectrum of Helium nuclei: the blue points and orange triangles are the results of measurements by AMS-02 \citep{AMS02 He-C-O} and Voyager \citep{VOYAGER} respectively. The lines are labelled as in Figure \ref{Fig 1: Proton}.}
	\label{Fig: Helium}
\end{figure}

\begin{figure}
	\includegraphics[width=0.5\textwidth]{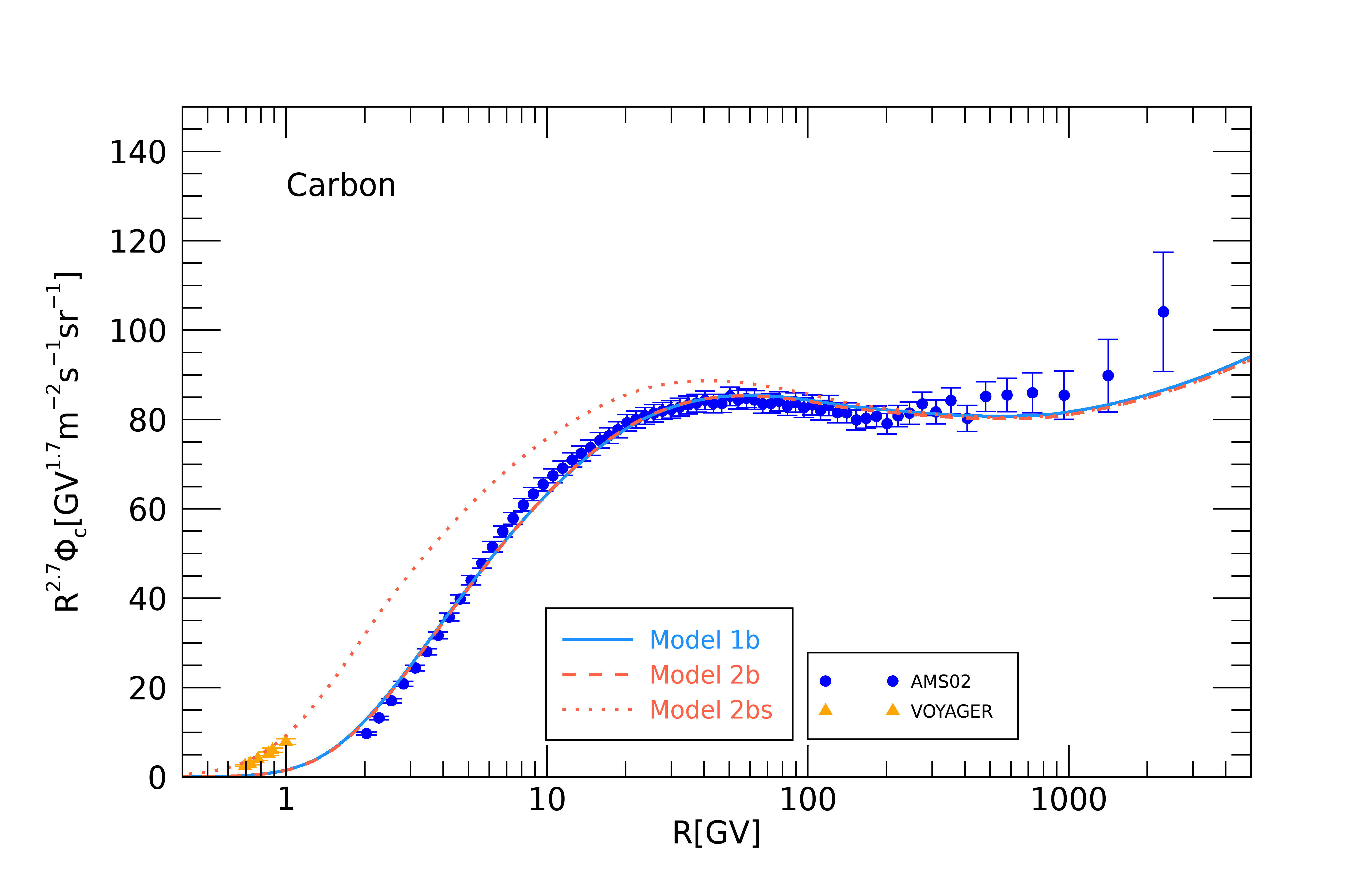}	
	\caption{Spectrum of Carbon nuclei: the blue points and orange triangles are the results of measurements by AMS-02 \citep{AMS02 He-C-O} and Voyager \citep{VOYAGER} respectively. The lines are labelled as in Figure \ref{Fig 1: Proton}.}
	\label{Fig: Carbon}
\end{figure}

\begin{figure}
	\includegraphics[width=0.5\textwidth]{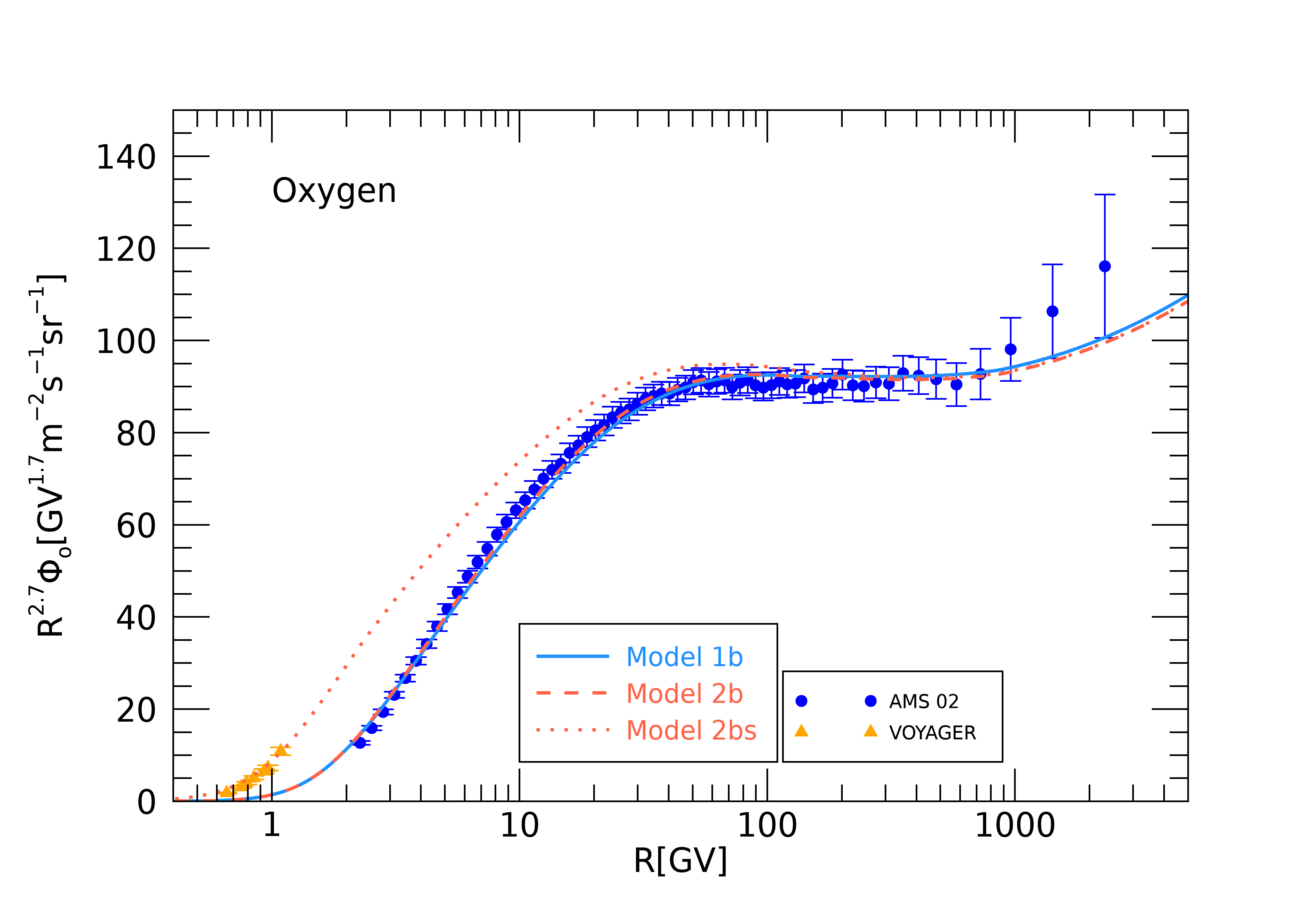}	
	\caption{Spectrum of Oxygen nuclei: the blue points and orange triangles are the results of measurements by AMS-02 \citep{AMS02 He-C-O} and Voyager \citep{VOYAGER} respectively. The lines are labelled as in Figure \ref{Fig 1: Proton}.}
	\label{Fig: Oxygen}
\end{figure}

\begin{figure}
	\includegraphics[width=0.5\textwidth]{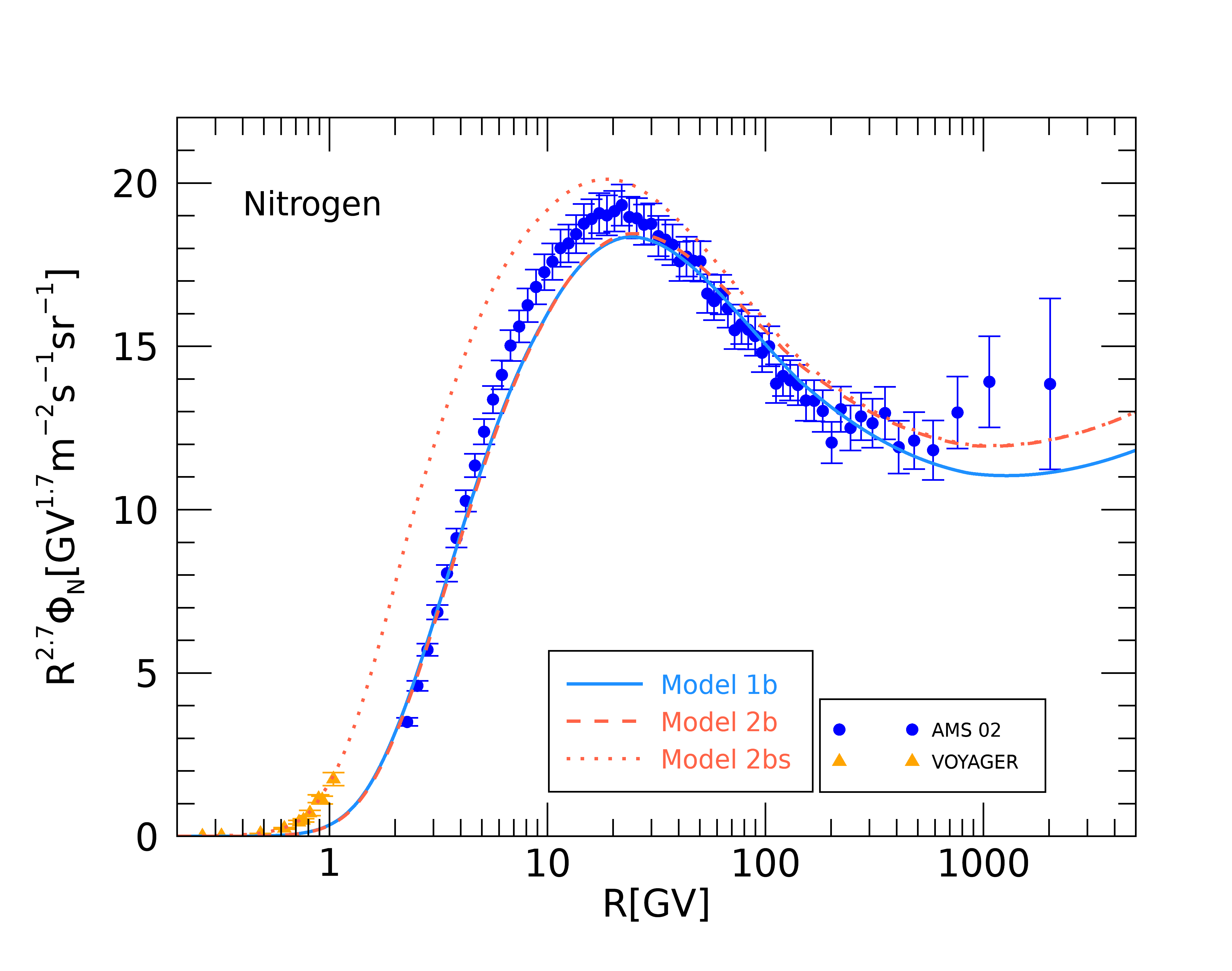}	
	\caption{Spectrum of Nitrogen nuclei: the blue points and orange triangles are the results of measurements by AMS-02 \citep{AMS02 He-C-O} and Voyager \citep{VOYAGER} respectively. The lines are labelled as in Figure \ref{Fig 1: Proton}.}
	\label{Fig: Nitrogen}
\end{figure}

All spectra clearly show the hardening at high energy induced by the shape of the diffusion coefficient, as found by \cite{Aloisio 2015}, thereby providing a good fit to the AMS-02 data. The spectrum of protons at low energies is dominated by the role of advection and, as visible in Fig.~\ref{Fig 1: Proton}, this allows for an excellent description of Voyager data. We conclude that the small discrepancy between model and data below $10$ GeV is to be attributed to a poor modelling of solar modulation. As expected, the role of reacceleration on the spectra of primary nuclei is negligible, as can be appreciated by comparing the solid and dashed lines in Fig.~\ref{Fig 1: Proton}. Similar considerations apply to the spectrum of Oxygen, shown in Fig.~\ref{Fig: Oxygen}. The observed fluxes of Helium (Fig.~\ref{Fig: Helium}) and Carbon (Fig.~\ref{Fig: Carbon}) contain a substantial secondary contribution as due to the production of $^{3}$He (which is added to the flux of $^{4}$He) and Carbon from spallation of heavier nuclei. Such secondary contributions suffer the action of reacceleration. Yet, the effects on the total flux of Helium and Carbon nuclei remain relatively small. 

The case of Nitrogen is more interesting in that its flux is dominated by secondary production at low energies and it is basically of primary origin at high energies. The flux is well described by the results of our calculations at all energies, and a sizeable contribution of reacceleration of secondary Nitrogen is clearly visible at high energy in Fig.~\ref{Fig: Nitrogen}. The small discrepancy at rigidity around $\lesssim 10$ GV is still much smaller than the uncertainty associated with the cross sections of Nitrogen production in spallation reactions. 



\begin{figure}
	\includegraphics[width=0.5\textwidth]{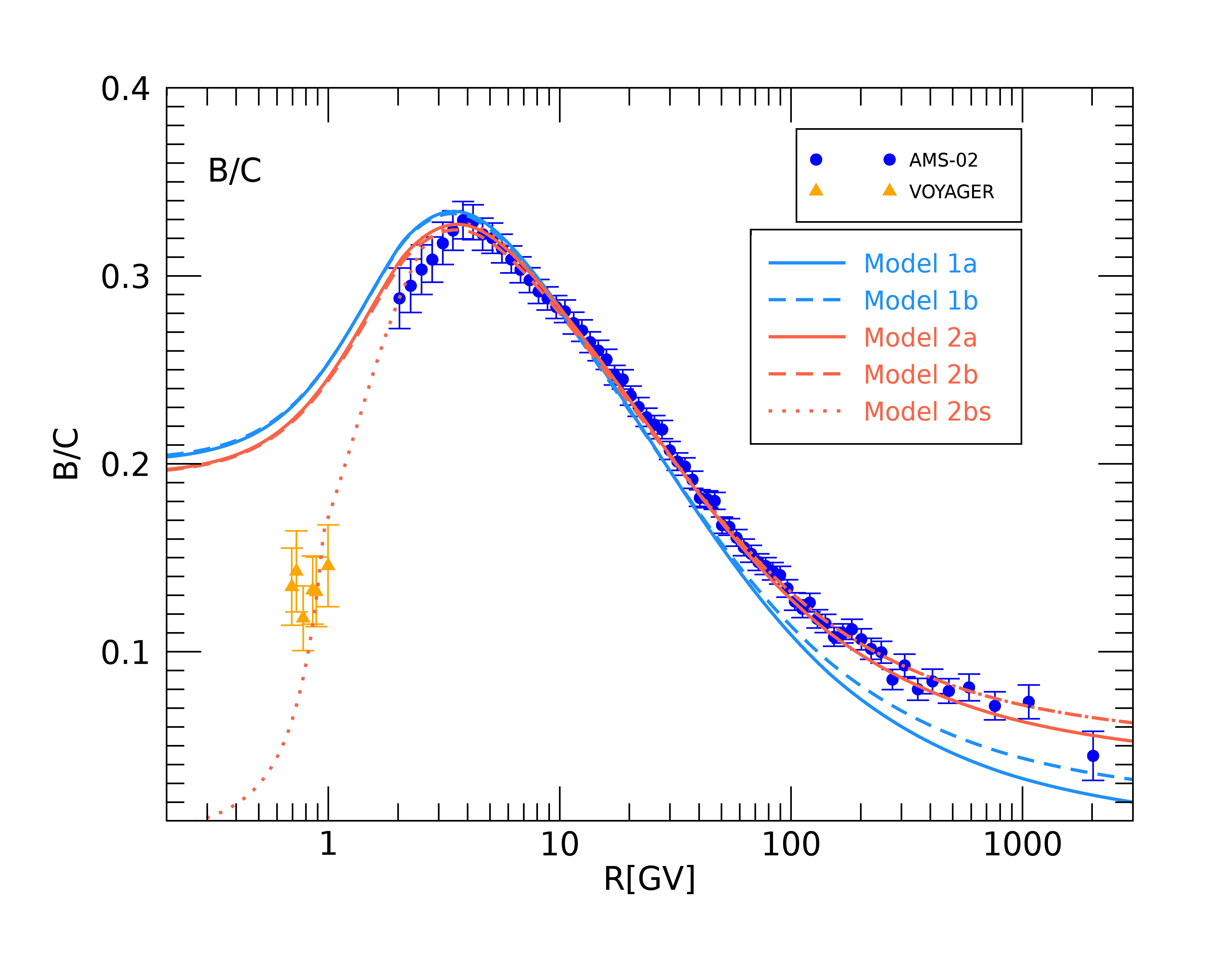}	
	\caption{Ratio of Boron over Carbon fluxes: the blue points and orange triangles are the results of measurements by AMS-02 \citep{AMS02 secondaries} and Voyager \citep{VOYAGER} respectively. The curves illustrate our results for the case without reacceleration  (Models 1-) and reacceleration (Models 2-). Dashed lines (Models -b) are obtained considering the additional grammage accumulated inside the sources, $X_{\rm src}$. Model 2bs refers to this latter case without inclusion of the effects of solar modulation, for direct comparison with Voyager data. See also Table \ref{tab:mod}.}
	\label{Fig: B/C}
\end{figure}

As already found by \cite{Blasi 2017}, the effect of reacceleration is much more prominent in the case of secondary nuclei. In particular the B/C data provide the best source of information on the grammage, which in turn is affected by the presence of reacceleration. The comparison between our predictions and the B/C data is shown in Fig.~\ref{Fig: B/C}. 
In the presence of reacceleration, the grammage at, say, 10 GV rigidity is reduced to $\sim 8.2$ g/cm$^2$, to be compared with $\sim 9.9$ g/cm$^2$ estimated in the absence of this phenomenon.

For the diffusion coefficient adopted in this work, the calculations carried out without accounting for reacceleration lead to a B/C ratio that fails to describe the data for $R\gtrsim 50$ GV (blue solid line in Fig.~\ref{Fig: B/C}), even in the case in which a source grammage is added (dashed line). On the other hand, including the reacceleration of Boron in random encounters with SNRs whose sizes are distributed according to the probability function discussed in \S~\ref{sec: src terms}, we obtain an excellent account for the B/C data up to the highest energies where measurements are available (red solid line in Fig.~\ref{Fig: B/C}). Adding a source grammage does not affect the results as much (red dashed line in the same figure). 

\begin{figure}
	\includegraphics[width=0.5\textwidth]{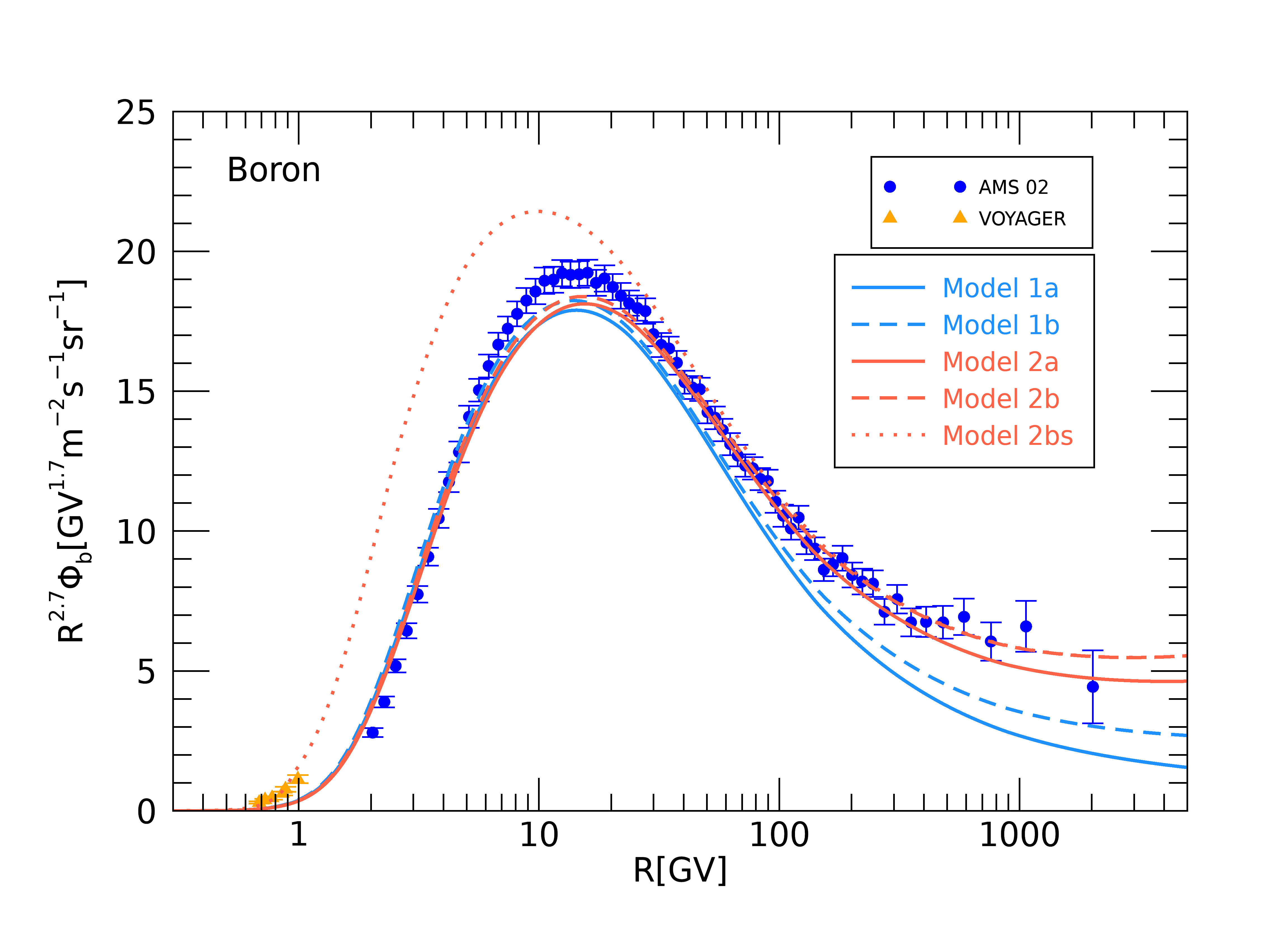}
	\includegraphics[width=0.5\textwidth]{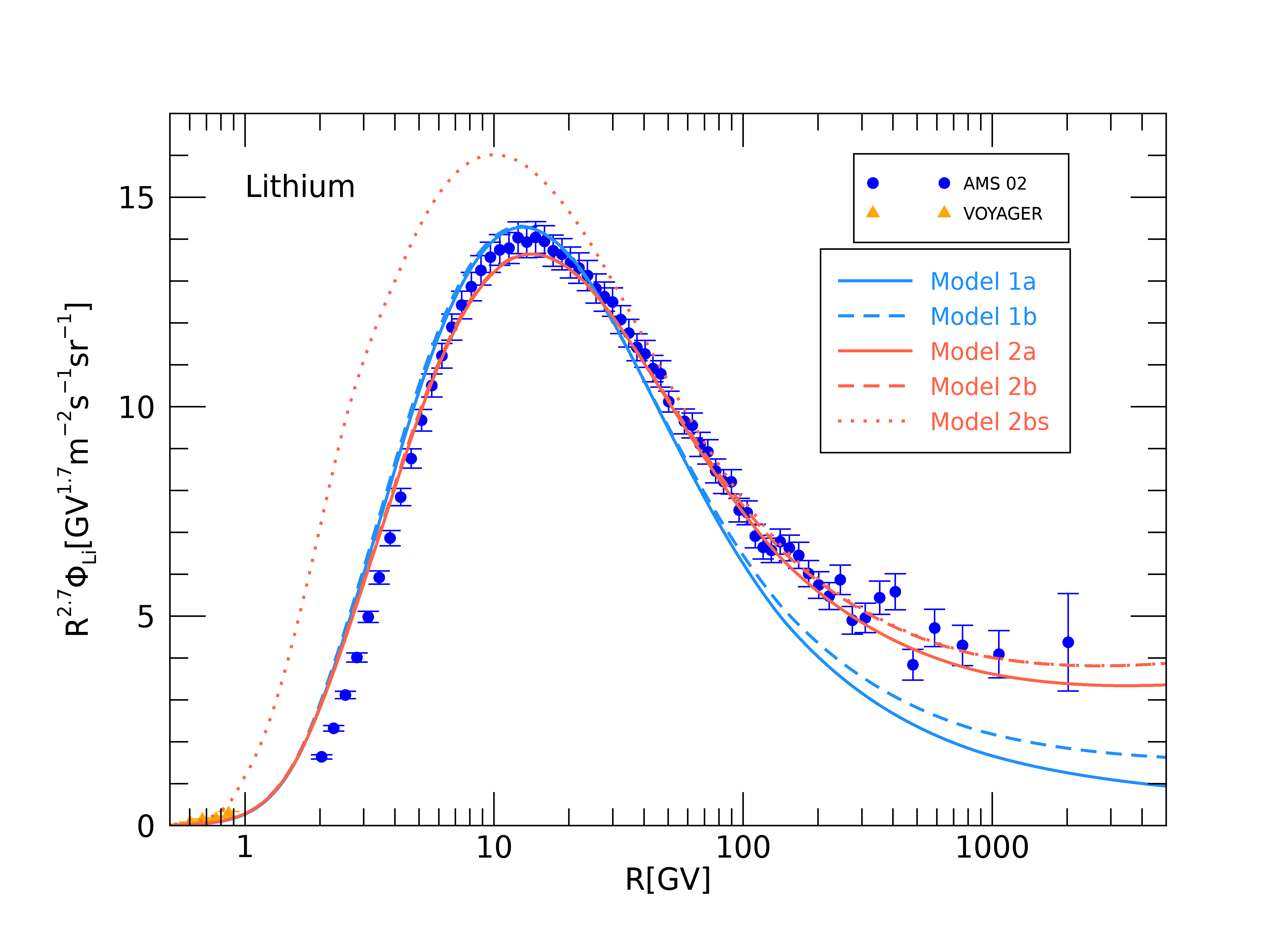}	
	\caption{Spectrum of B (upper panel) and Li nuclei (lower panel): the blue points and orange triangles are the results of measurements by AMS-02 \citep{AMS02 secondaries} and Voyager \citep{VOYAGER} respectively. The lines are labelled as in Figure \ref{Fig: B/C}. Dashed lines are obtained considering the additional source term deriving from spallation processes suffered by primaries (C+O) before escaping from the source.}
	\label{Fig: B}
\end{figure}

For completeness we also show the spectrum of Boron and Lithium, both secondary products of spallation reactions (Fig.~\ref{Fig: B}). The effect of reacceleration is again very clear from the comparison between the blue (Model 1) and red (Model 2) curves, while the source grammage has a lesser impact. The role of reacceleration is clear: in its absence, the spectra of secondaries would scale as $E_k^{-s+2-2\delta}$, if $\delta$ is the slope of the diffusion coefficient, while the occasional encounters of secondary nuclei with SNR shocks create a subdominant component with a harder spectrum, $\propto E_k^{-s+2-\delta}$, which becomes important at high energies. We stress again that the unmodulated B/C ratio and the unmodulated spectra of Boron and Lithium are all in good agreement with Voyager data at low energies. The slight underprediction of the Boron peak flux is not unexpected and due to the fact that our calculations do not include the production of $^{10}B$ from the decay of $^{10}Be$.  

%
\begin{figure}
	\includegraphics[width=0.5\textwidth]{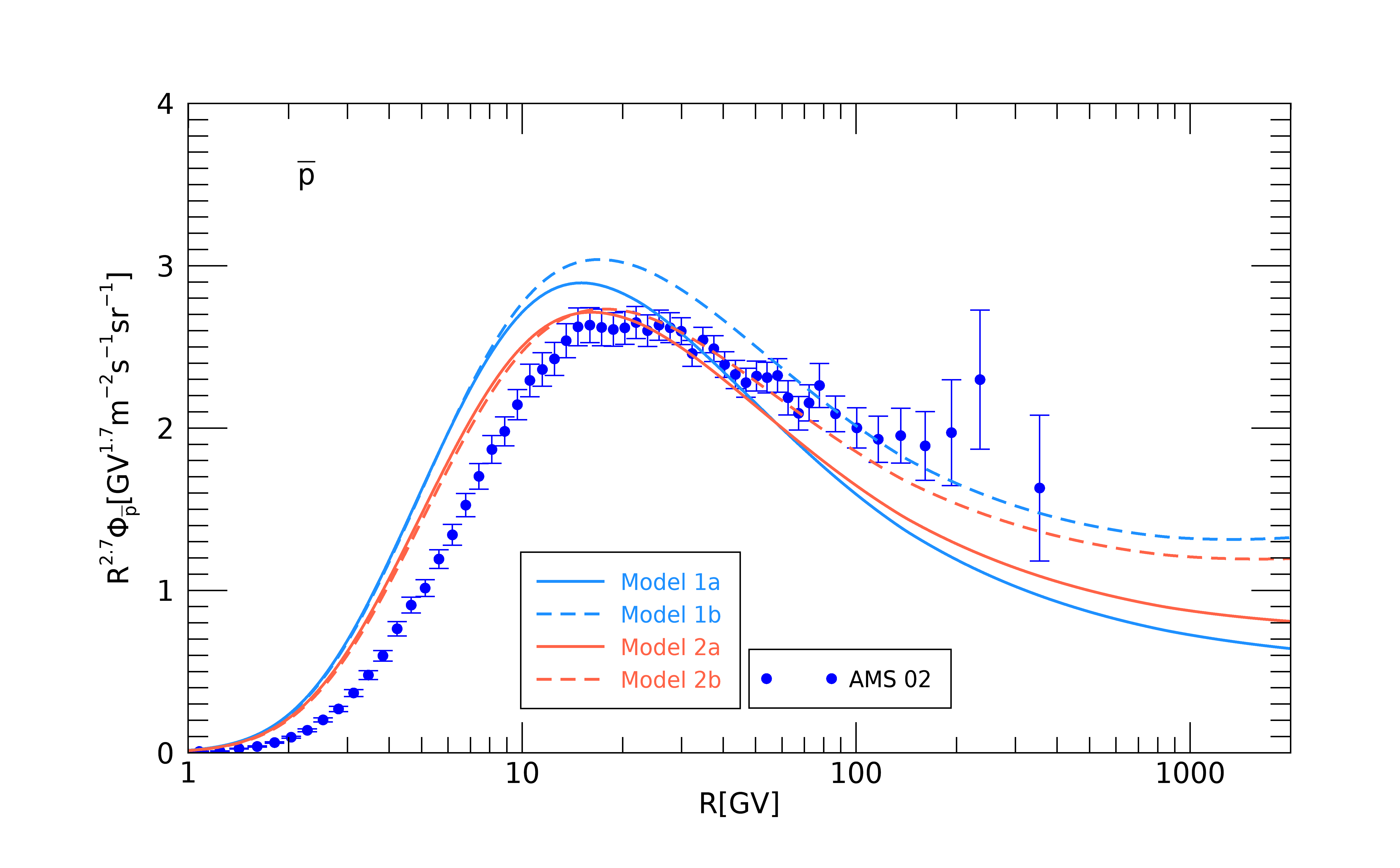}	
	\caption{Spectrum of antiprotons: the blue points are the results of measurements by AMS-02 \citep{pbar}. The lines are labelled as in Figure \ref{Fig: B/C}. Dashed lines are obtained considering the additional source term in the antiprotons flux deriving from interactions suffered by primaries (p+He) before the escape from the source (Term in Eq. \ref{eq:pbarinjsrc}).}
	\label{Fig: Pbar}
\end{figure}
%

Finally we consider antiprotons as secondary products of CR interactions in the ISM. The $\bar{p}$ spectrum is shown in Fig.~\ref{Fig: Pbar} for the same models discussed above (except that there are no Voyager data on antiprotons, hence we do not show the unmodulated $\bar{p}$ spectrum). The comparison between the two models, with and without reacceleration, seems to show a less pronounced effect of reacceleration for $\bar{p}$. This happens because reacceleration is less effective when the spectrum of the secondary particles is harder. The spectrum of $\bar{p}$ is harder than that of Boron and Lithium for two reasons: 1) while B and Li are produced at the same energy per nucleon of the parent nucleus, antiprotons of given energy are produced in inelastic collisions of primaries with energy $\sim 10$ times larger, on average; for $E_{\bar p}\gtrsim 10$ GeV, this corresponds to primary protons with energy above the break where the spectrum of protons is harder,  $\propto E_k^{-s+\beta+2}$. 2) The cross section for $\bar{p}$ production is a growing function of energy, say $\sigma_{\bar p}\propto E^{\alpha}$. Hence, the equilibrium spectrum of antiprotons is $\sim E_k^{-s+\beta+2-2\delta+\alpha}$, rather than simply $\sim E_k^{-s+2-2\delta}$, as in the case of B and Li. These two effects make the antiproton spectrum harder than that of Boron, and the effect of reacceleration correspondingly weaker, in the sense that reacceleration becomes important at higher energies. On the other hand, the source grammage plays an important role on the overall $\bar{p}$ spectrum, as one can see from a comparison between the blue and red dashed lines in Fig.~\ref{Fig: Pbar}. Reacceleration remains important in terms of implying a smaller normalization of the grammage, which implies that the $\bar{p}$ data are best described when reacceleration and source grammage are both included (red dashed line) in Fig.~\ref{Fig: Pbar}. In this picture, the behaviour of antiprotons does not appear to require alternative descriptions of the CR transport, as postulated by \cite{lipari}, although such models are still worth being investigated for the reasons discussed earlier in this article. At rigidities lower than $ \sim 10$ GV we are not able to recover a good agreement between data and antiproton flux. Below that rigidity the solar wind modulation plays an important role and should probably be treated more carefully, considering the dependence on particle charge, polarity of the Sun magnetic field and the specific times when data have been collected.

\section{Conclusions} 
\label{sec: conclusion}

Research in CR physics has been impacted quite substantially by the AMS-02 measurements, not only because of the numerous unexpected results, but also because the high precision of the collected data, up to TeV energies, has forced the community to reconsider the subtleties of propagation. On one hand, we now need to worry about effects that were known to exist but were neglected because their impact on observables used to be smaller than the error bars. On the other hand, we are forced to think of phenomena, either conventional or unconventional, that we might have neglected. An instance of the former type of effects is the source grammage, which to some extent has been known to exist but was typically ignored because of its small value. An instance of the latter type of phenomena, instead, is the reacceleration of secondary CR products due to occasional crossing of SNR shocks \citep{Blasi 2017}. In this article we discussed these two effects and showed that both of them have a potentially important impact on the spectrum of secondary nuclei such as Boron and Lithium, and also on partially secondary nuclei, such as Nitrogen. In addition, antiprotons, produced as a result of inelastic collisions of CR protons and Helium with the ISM, are also deeply affected by the source grammage and reacceleration processes, which make their high energy spectrum harder than naively expected. 

These findings acquire an even more prominent importance whenever the AMS-02 data are considered as indicative of some radically new picture of CR transport, as advocated by e.g. \cite{Katz2010,Blum2013,lipari}, mainly based on the anomalous behaviour of CR positrons and antiprotons. While for positrons viable alternative explanations of the rising fraction have been proposed, for instance associated with mature pulsars \citep{hooper,pulsars}, antiprotons immediately appeared to be more challenging. As pointed out by \cite{lipari}, the fact that the spectral shapes of protons, antiprotons and positrons are very similar, contrary to what expected based on the standard model of CR transport, seems to suggest that both positrons and antiprotons are simply secondary products of primary CRs, and it is the conventional scenario of CR origin and transport that must be revised. The alternative models so far proposed, however, come with new unanswered questions: they imply the lack of radiative energy losses for leptons, very different source spectra for electrons and protons, and a very steep injection for protons, for which the commonly assumed $E^{-2.2}$ spectrum is already problematic to account for (see e.g.\cite{Cardillo et al 15}). It appears appropriate, then, to make sure that the {\it anomalies} highlighted by the new data cannot be accounted for within the standard scenario of CR propagation, before abandoning it. In this work we have made a first step towards including effects that must be there at some level, and so far were simply neglected.

An important ingredient of our entire calculation, which deserves further discussion, is the energy dependence of the diffusion coefficient. We adopted the diffusion coefficient found by \cite{Aloisio 2015}, as a result of a non-linear calculation of CR propagation, where the spectra of the most abundant CR nuclei were reproduced with the diffusion coefficient deriving from the spectrum of waves self-consistently computed. Such a diffusion coefficient has a break at a rigidity of $\sim 300$ GV, as a result of the transition between self-generated waves, with a steep spectrum, at lower rigidities, and a Kolmogorov turbulence spectrum at higher rigidities. This kind of rigidity dependence of the diffusion coefficient is exactly what is needed to explain the hardening observed in the spectra of virtually all elements in CRs \citep{cream1, Pamela break, AMS02 Protons, AMS02 Helium, cream2} . The recent finding \cite[]{AMS02 secondaries} that such hardening is more pronounced for secondary CRs than for primaries strongly supports the idea that it originates from a change in the particle transport regime, rather than from some peculiarity of the acceleration process. 

While fitting the spectra of primary and secondary nuclei with a diffusion coefficient whose energy dependence is free, both below and above the break, is certainly doable, as long as one allows for a different injection spectrum from protons, He and heavier nuclei  \citep{Evoli}
, the diffusion coefficient we adopt in this article does not allow to reproduce the $B/C$ ratio at rigidities $R\gtrsim 50$ GV, a result that was already pointed out by \cite{Aloisio 2015}. We showed however that the spectra of both primary and secondary nuclei and the spectrum of antiprotons can all be accounted for when the effects of reacceleration and source grammage are included. Moreover, the injection spectra of protons and CNO are very close to each other (4.2 versus 4.18) and larger than the required slope for Helium (4.12). In the absence of reacceleration, the slope of the injection spectrum of nuclei heavier than Helium is $\sim 0.05$ harder than that of protons, while Helium requires an even harder spectrum (harder by $\sim 0.1$ with respect to that of protons) \citep{Evoli}. The steep spectra of secondary nuclei, compared with the equilibrium spectrum of antiprotons, make the effect of reacceleration more important for secondary nuclei, while the spectrum of antiprotons is more affected by the presence of a source grammage. It should be said that from the theoretical point of view, reacceleration is somewhat more general, in that it takes place independent of the conditions in which a SNR shock propagates (it only depends on the shock velocity and size), while the source grammage is accumulated only when the SN explosion occurs in a relatively dense medium ($\sim 1\rm cm^{-3}$). 

The purpose of the work presented here was not that of obtaining a detailed fit to the data. The required scanning of the parameter space is particularly challenging, given the recursive nature of the solution procedure in the presence of reacceleration. Our purpose was rather that of clarifying how known astrophysical phenomena may play a crucial role in explaining the data, once the experimental error bars become at the percent level, as it is the case for AMS-02 measurements. Since the need to revisit the pillars of the generally accepted model for the origin of CRs is often invoked, such efforts should always be accompanied by an attempt to establish that known phenomena play a marginal role.

\section*{Acknowledgements}
We are grateful to Carmelo Evoli and Michael Korsmeier for help with the cross sections involved in spallation and antiproton production processes respectively. This work was partially funded through Grants ASI/INAF n. 2017-14-H.O and SKA-CTA-INAF 2016.

\appendix
\appendix
\section{Numerical approximation for the propagation integral}
The semi-analytical formalism used in this article requires the numerical evaluation of the integral in Eq.~\ref{eq:sluI0ek} to find the solution for the CR distribution function. This kind of integral often appears in the solution of the transport equation when a damping term is present. We present here a numerical technique that allows us to calculate it quickly and with high accuracy. A general propagation integral written in the form
\begin{equation} \label{eq:A1}
  F(x)= \int_{x_0}^{x} f(x') \, e^{-\int_{x'}^x \alpha(y) dy} dx'
\end{equation}
describes how the signal $f$ produced in $x'$ propagates to the position $x$ when a damping term (or absorption term), $\alpha$, is present (note that we assume $\alpha > 0$). Due to the presence of the exponential term, one has to pay special attention when the propagation integral (Eq.~\ref{eq:A1}) is solved numerically on a grid. The exponential term peaks at $x=x'$ with a typical width of $\sim 1/\alpha$, which implies that one should choose the step-size of the grid $\delta\ll 1/\alpha$. The difficulty in our case is that at each position $x$, $\alpha$ can vary by orders of magnitude because it depends on the particle kinetic energy $E_k$ (Eq.~\ref{eq:sluI0ek}). Hence we are forced to choose a very fine grid, with a step-size that is extremely small compared to the spatial scale of the problem. As a result the numerical calculation turns out to be extremely time consuming. To overcome this difficulty, when we perform the numerical integration on a grid, we start from the value of $F$ in a position $x$ and we calculate $F$ in the next position $x+\delta$ using the following identity
\begin{equation} \label{eq:A2}
F(x+\delta)= e^{-\bar \alpha \delta} \left[ F(x) + \int_{x}^{x+\delta} f(x') \, e^{-\int_{x'}^{x} \alpha(y) dy} dx' \right] \,,
\end{equation}
where $\bar \alpha \equiv \delta^{-1} \int_{x}^{x+\delta} \alpha(y) dy$ is the average value of $\alpha$ in the range $[x,x+\delta]$. In this same range we approximate $f(x)$ with a linear function, hence the first derivative is $f' \simeq (f_2-f_1)/\delta$, where $f_1= f(x)$ and $f_2=f(x+\delta)$. We also approximate the argument of the exponential with a linear function, hence $\int_{x'}^{x} \alpha(y) dy \simeq \bar \alpha (x-x')$.  Using these two approximations, the integral in Eq.~\ref{eq:A2} can be solved by parts, leading to the following expression:
\begin{equation} \label{eq:A3}
\begin{split}
F(x+\delta) &= e^{-\bar \alpha \delta} \left[ F(x) + \frac{f_2-f_1 (1+\bar \alpha \delta)}{{\bar \alpha}^2 \delta} \right]+ \\
&+ \frac{f_1-f_2 (1- \bar \alpha \delta)}{{\bar \alpha}^2 \delta}
+ \frac{1+ e^{-\bar \alpha \delta}}{\bar \alpha^2} \mathcal{O}(\delta^3 f'') \,.
\end{split}
\end{equation}
Here the error $\mathcal{O}()$ is proportional to $\delta^3$ times the value of the function second derivative somewhere in the interval of integration. Now the condition to fulfil in order to have a negligible error is much less restrictive and reads $\delta\ll (\bar \alpha/ f'')^{1/4}$.
One can easily get a better approximation using higher order approximations for $f$ and $\alpha$. Nevertheless, for our purposes, Eq.~\ref{eq:A3} is sufficient to reach an acceptable accuracy. Finally we note that the expression Eq.~\ref{eq:A3} returns the correct result when the damping term dominates with respect to the propagation, namely when $\bar \alpha \delta \gg 1$. This limit applies when the damping is so strong that the function $F$ is only determined by the source term $f$. In this limit, the exponent in Eq.~\ref{eq:A1} behaves like a Dirac $\delta$-function and we have $F(x) \approx f(x)/\alpha(x)$.

When $\bar{\alpha} \delta \ll 1$, Eq.~\ref{eq:A3} is not accurate. A better solution is obtained by Taylor expanding the exponential up to the third order, namely
\begin{equation} \label{eq:A4}
e^{-\bar{\alpha} \delta} = 1- \bar{\alpha} \delta + \frac{(\bar{\alpha} \delta)^2}{2} - \frac{(\bar{\alpha} \delta)^3}{6} +  \mathcal{O}(\bar{\alpha} \delta)^4) \,.
\end{equation}
Using this expansion in Eq.~\ref{eq:A2} and integrating by parts, we obtain:
\begin{equation} \label{eq:A5}
F(x+\delta) = F(x) e^{-\bar \alpha \delta}  + \frac{\delta}{6} 
\left[ f_2 (3-\bar \alpha \delta) + f_1 (3- 2\bar \alpha \delta + {\bar \alpha}^2 \delta^2)\right]\,.
\end{equation}
Finally we also write the solution for the case when the known boundary condition of the problem is known at the upper end of the integration interval. In this case Eq.~\ref{eq:A1} reads:
\begin{equation} \label{eq:A6}
F(x) = \int_{x}^{x_{\max}} f(x') \, e^{-\int_{x}^{x'} \alpha(y) dy} dx' \,.
\end{equation}
Following the same reasoning as above, we find a recursive formula similar to Eq.~\ref{eq:A3} where $f_1$ replaces $f_2$ and vice-versa:
\begin{equation} \label{eq:A7}
\begin{split}
F(x-\delta)& = e^{-\bar \alpha \delta} \left[ F(x) + \frac{f_1 - f_2(1+\bar \alpha \delta) }{{\bar \alpha}^2 \delta} \right] +\\
&+ \frac{f_2 - f_1 (1- \bar \alpha \delta) }{{\bar \alpha}^2 \delta}
+ \frac{1+ e^{-\bar \alpha \delta}}{\bar \alpha^2} \mathcal{O}(\delta^3 f'') \,.
\end{split}
\end{equation}




\bsp	
\label{lastpage}
\end{document}